\documentclass{amsart}
\usepackage{graphicx}
\usepackage{amssymb}
\usepackage{latexsym}
\usepackage{amsmath}
\usepackage{bbm}
\usepackage{float}

\usepackage{scalerel,stackengine}
\stackMath
\newcommand\reallywidecheck[1]{%
\savestack{\tmpbox}{\stretchto{%
  \scaleto{%
    \scalerel*[\widthof{\ensuremath{#1}}]{\kern-.6pt\bigwedge\kern-.6pt}%
    {\rule[-\textheight/2]{1ex}{\textheight}}
  }{\textheight}%
}{0.5ex}}%
\stackon[1pt]{#1}{\scalebox{-1}{\tmpbox}}%
}
\usepackage{color}
\usepackage{xcolor}
\usepackage[breaklinks,colorlinks,backref]{hyperref}
\hypersetup{
    colorlinks, 
    linktoc=all, 
    linkcolor=red,  
  citecolor=hyptxt,
  urlcolor=blue}
  \hypersetup{
  citebordercolor=,
  filebordercolor=green,
  linkbordercolor=blue
}
\definecolor{hyptxt}{rgb}{0.7, 0.4, 0.9}
\usepackage{bibtopic}

\newcommand{\vsi}{\varsigma}
\newcommand{\vsd}{\varsigma_{_{\text{\scriptsize dS}}}}
\newcommand{\vsa}{\varsigma_{\text{\scriptsize AdS}}}
\newcommand{\be}{\beta}
\newcommand{\ga}{\gamma}

\newcommand{\al}{\alpha}

\newcommand{\lga}{\longrightarrow}

\newcommand\bei{\begin{itemize}}
\newcommand\ei{\end{itemize}}

\def\lg{\langle}
\def\rg{\rangle}
\def\ud{\mathrm{d}}

\def\R{\mathbb{R}}

\def\C{\mathbb{C}}

\def\SD{\rtimes}
\def\ii{\mathrm{i}}

\begin{document}

\markboth{G.~Cohen-Tannoudji  \& J.-P.~Gazeau }
{Quantum fields with no Planck constant}

\title[Dark matter as AdS  gluon BEC]{Cold Dark Matter: A Gluonic Bose--Einstein Condensate in Anti-de Sitter Space Time}

\author[Cohen-Tannoudji \& Gazeau]{Gilles Cohen-Tannoudji$^a$ and Jean-Pierre Gazeau$^b$}

\address{$^a$ Laboratoire de recherche sur les sciences de la mati\`ere, LARSIM CEA, Universit\'e Paris-Saclay, F-91190 Saint-Aubin, France\\                                       
Gilles.Cohen-Tannoudji@cea.fr }

\address{$^b$ Universit\'e de Paris, CNRS, Astroparticule et Cosmologie, F-75013 Paris, France  
\\ gazeau@apc.in2p3.fr}

\begin{abstract}In the same way as the realization of some of the famous gedanken experiments imagined by the founding fathers of quantum mechanics has recently led to the current renewal of the interpretation of quantum physics, it seems that the most recent progresses of observational astrophysics can be interpreted as the realization of some cosmological gedanken experiments such as the removal from the universe of the whole visible matter or the cosmic time travel leading to a new cosmological standard model. This standard model involves two dark components of the universe, dark energy and dark matter. Whereas dark energy is usually associated with the  cosmological constant, we propose to explain dark matter as a pure QCD effect, 
namely a gluonic Bose Einstein condensate, following  the transition from the quark gluon plasma phase to the colorless hadronic phase.
 Our approach not only allows us to assume  a ratio Dark/Visible  equal to 11/2 but also provides gluons (and di-gluons, viewed as quasi-particles)  with an extra mass of vibrational nature.  Such an interpretation would comfort the idea that, apart from the violation of the matter/antimatter symmetry satisfying the Sakharov’s conditions, the reconciliation of particle physics and cosmology  needs not the recourse to any ad hoc fields, particles or hidden variables.
\end{abstract}


\maketitle


\tableofcontents

\section{Introduction}

   The \emph{new cosmological standard model} model involves two dark components of the universe, dark energy \cite{peebles03} and dark matter\\ \cite{arbey21}. Whereas dark energy is commonly associated with the cosmological constant, both of us, Gilles Cohen-Tannoudji \cite{GCT1} and \cite{GCT2}, and Jean-Pierre Gazeau \cite{gazeau1-20} have independently tried to address the challenging issue of the dark matter component in the cosmological energy density. 
 
The approach of GC-T \cite{GCT1} and \cite{GCT2} aimed at interpreting dark matter as a component of the cosmological energy density, which, together with dark energy, would constitute the \emph{world matter}, namely what, according to de Sitter, must be added to the visible matter in order, for a cosmological theory to obey the principle of the relativity of inertia. On the other hand, the interpretation by J-PG in \cite{gazeau1-20} in terms of a pure vibration energy due to positive curvature was partially based on mass formulae in terms of energy and spin in de Sitter and/or Anti-de Sitter spacetimes, that are established in the quantum context with the reasonable assumption  that the proper mass of an elementary system (in the Wigner \cite{wigner39,bargwig48} sense) is independent of the space-time metric.

In the present paper, we explain how our two approaches are complementary in proposing the value $11/2=5.5$  for  the   ratio Dark/Visible (the observed one is currently estimated to be $27/5=5.2$) and  interpreting (cold) dark matter  as a gluonic Bose Einstein Condensate (BEC) that is a relic of the quark period. These proposals  were already present in \cite{GCT2} but the BEC interpretation is here given a firmer basis thanks to the quantum features of elementary systems in Anti-de Sitter spacetime whose the description was given in  \cite{gazeau1-20}.

In Section \ref{gedank}, we   review the history of the new cosmological standard model, known as $\Lambda$CDM,  from the Einstein-de Sitter debate at the onset of modern cosmology to its current assets. We recall that $\Lambda$ stands for the cosmological constant and CDM for Cold Dark Matter. 

Section \ref{dSAdSL} is devoted to the key concept of a \emph{(normal- or anti-) de Sitter comoving world matter density}     with which we intend to fill the gaps between our two approaches.  

In Section \ref{HPCOSMO} we conjecture, as a merging of our two approaches, that the cold dark matter, identified with the gravitational potential induced by matter, is affected by a Bose-Einstein condensation mechanism. 

Finally, in Section \ref{disc}, we  compare our approach with other schemes that also assume a Bose-Einstein mechanism, but involving unknown particles.

\section{The new standard model of cosmology, the performance of the critical cosmological gedanken experiment and its qualitative results}
\label{gedank}
\subsection{The Einstein de Sitter debate}

Our common starting point is the history of the debate that was raised between Einstein \cite{einstein17}  and de Sitter \cite{desitter17}, at the onset of modern cosmology. This debate was about a critical cosmological gedanken experiment, the one which would consist of ``removing all the visible matter from the universe'' in order to decide whether or not an isolated particle, acting as a test body would have inertia. The answer to this question refers to what is known as the \emph{Mach’s principle} or what Einstein and de Sitter named \emph{the principle of the relativity of inertia}. It is summarized in the following quoted statement by de Sitter \cite{desitter17}, in which he introduces the concept of \emph{world matter}:

\begin{quote}
\emph{To the question: If all matter is supposed not to exist, with the exception of one material point which is to be used as a test-body, has then this test-body inertia or not? The school of Mach requires the answer No. Our experience however decidedly gives the answer Yes, if by `all matter' is meant all ordinary physical matter: stars, nebulae, clusters, etc. The followers of Mach are thus compelled to assume the existence of still more matter: the \emph{world-matter}.} \footnote{The proper mass is predicted by special relativity if we adopt Wigner point of view of elementary system \cite{wigner39,bargwig48}. Its existence results from the symmetry of empty Minkowski, de Sitter, and Anti-de Sitter space-times as being one of the two invariants (the other one being the spin) of the representations of their respective kinematical groups.  This point is developed  in Subsection \ref{WMdSAdS}.}  	
\end{quote}
	
	The debate of Einstein and de Sitter concerned three possible cosmological models: the first one was the first Einstein’s closed, static model consisting of large masses sent at spatial infinity, a model that Einstein abandoned following the criticism of de Sitter; in the second model which de Sitter calls the `system A', Einstein re-introduces the `cosmological term', involving the cosmological constant that he had ignored in previous attempts, leading to a sort of repulsive force (negative pressure) preventing the universe from collapsing under its own gravitation, and which de Sitter assimilates to a world matter insuring the validity of the postulate of the relativity of inertia; and the third one, the `system B' according to de Sitter consists of a universe that is empty except for the cosmological term. About these last two models, de Sitter summarizes the debate in a postscript added to Ref. \cite{desitter17} and quoted here:

\begin{quote}	
\emph{Prof. Einstein, to whom I had communicated the principal contents of this paper, writes `to my opinion, that it would be possible to think of a universe without matter is unsatisfactory. On the contrary the field $g^{\mu \nu}$ \emph{must be determined by matter, without which it cannot exist} [underlined by de Sitter]. This is the core of what I mean by the postulate of the relativity of inertia'. He therefore postulates what I called above the logical impossibility of supposing matter not to exist. I can call this the ``material postulate'' of the relativity of inertia. This can only be satisfied by choosing the system A, with its world-matter, i.e. by introducing the constant $\lambda$, and assigning to the time a separate position amongst the four coordinates. On the other hand, we have the `mathematical postulate' of the relativity of inertia, i.e. the postulate that the $g^{\mu \nu}$ shall be invariant at infinity. This postulate, which, as has already been pointed out above, has no physical meaning, makes no mention of matter. It can be satisfied by choosing the system B without a world-matter, and with complete relativity of the time. But here also we need the constant $\lambda$.  \underline{The introduction of this constant can only be avoided}  \underline{by abandoning the postulate of the relativity of inertia altogether}} [underlined by us].
\end{quote}

By revisiting the Einstein-de Sitter debate about the concept of inertia, one can notice a perplexing irresolution concerning the position  of that constant $\lambda$ or $\Lambda$ in the left-hand side (as a fundamental constant) or in the right-hand side (as a phenomenological world matter term) in the Einstein equation \cite{gaznov11}. This is also the insistent ``little music'' pervading the content of the present contribution. 
  
\subsection{The performance of the gedanken experiment with $\Lambda$CDM}
The more and more precise measurements of the cosmic microwave background (CMB) radiation by the COBE, WMAP, and Planck experiments allowed the performance of the above mentioned gedanken experiment leading to the assets of $\Lambda$CDM,  the new standard model of cosmology, namely:
\begin{itemize}
  \item The rediscovery of the cosmological constant that, as mentioned above, is essential for the validity of the foundational principle of the relativity of inertia.
  \item The replacement of the big bang singularity which prevented any causal description of the early universe by an inflation mechanism that remains conjectural but can explain quantitatively the primordial fluctuations observed in the CMB.
  \item The discovery of two non-visible components of the cosmological energy density, which together amount to about 95\% of the full content of the universe, the \emph{dark energy} that is commonly associated with the cosmological constant, and the \emph{dark matter} which raises the theoretical questions some of which are addressed in the present paper.
\end{itemize}

 \section{A possible kinematics in quantum cosmology: desitterian/anti-desitterian comoving world-matter densities}
 \label{dSAdSL}
 In this section we first remind the Friedmann-Lemaitre model  involving density and pressure of the material content of the universe before describing the consequences on the kinematic symmetry of space time if one decides to view pressure as an effective curvature.   
 \subsection{A reminder about the cosmological formalism ($c=1$)}
 In an isotropic and homogeneous cosmology, the solution of the Einstein’s equation 
 \begin{equation}
\label{eineq}
\mathsf{R}_{\mu\nu} - \frac{1}{2}g_{\mu\nu}\mathsf{R}= 8\pi G_N T_{\mu\nu} +\Lambda g_{\mu\nu}
\end{equation}
is the Robertson metric
\begin{equation}
\label{robmet}
\ud s^2 = \ud t^2 - R^2(t)\left(\frac{\ud r^2}{1-kr^2} + r^2(\ud \theta^2 + \sin^2\theta \,\ud\phi^2\right)
\end{equation}
depending on the time-dependent radius of the universe $R(t)$ and a curvature index $k$. The coordinate $r$  is dimensionless; the dimension is carried by $R(t)$, which is the cosmological scale factor which determines proper distances in terms of the comoving coordinates.  These quantities obey the Friedmann-Lema\^{\i}tre equations of a perfect fluid with which is phenomenologically modeled the material content of the universe.
\begin{equation}
\label{friedlem1}
{{H}^{2}}\equiv {{\left( \frac{{\dot{R}}}{R} \right)}^{2}}=\frac{8\pi {{G}_{N}}\rho }{3}-\frac{k}{{{R}^{2}}}+\frac{\Lambda }{3}
\end{equation}
\begin{equation}
\label{friedlem2}
\frac{{\ddot{R}}}{R}=\frac{\Lambda }{3}-\frac{4\pi {{G}_{N}}}{3}\left( \rho +3P \right)\,,
\end{equation}
and a third equation expressing energy conservation,
\begin{equation}
\label{friedlem3}
\dot{\rho }=-3H\left( \rho +P \right)\,.
\end{equation}
In these equations, the density  $\rho$ and isotropic pressure $P$ express the stress energy momentum of the perfect fluid:
\begin{equation}
\label{encons}
{{T}_{\mu \nu }}=-P{{g}_{\mu \nu }}+\left( P+\rho  \right){{u}_{\mu }}{{u}_{\nu }}\,.
\end{equation}
The cosmological term is taken to the right-hand side of the Einstein’s equation and may be interpreted as a contribution to the stress energy tensor that reduces to minus the pressure multiplying the metric $g_{\mu\nu}$  (the density and the pressure sum to zero). This represents a quantitative expression of what de Sitter called a world matter in his debate with Einstein. Let us note that according to the sign of the pressure one can talk of a de Sitter world matter ($\Lambda$ positive, pressure negative) or an anti-de Sitter world matter ($\Lambda$ negative, pressure positive).

\subsection{Dark matter as an anti-de Sitter world matter}
\label{DMADSWM}
In the measurement of the CMB radiation it is crucial to get rid of the light that is emitted in the foreground in order to obtain the original map of the CMB radiation. This can be done using known technics, but once this is done, one is faced with the problem of the gravitational lensing possibly distorting the path of light between its emission and its arrival at the detector. To solve this problem, it has been possible to use a technique that has been already used to get information about the dark matter present in some very heavy super clusters of galaxies: such a cluster may induce a gravitational lensing potential distorting (and possibly multiplying) the image of a galaxy situated far behind the cluster; correlating the distorted observations one has been able to produce a map of the dark matter present in the cluster or in its halo that induces the lensing. The success of this technique has been considered as a proof of the presence of dark matter at extra galactic scales. Using this technique for the full sky distribution of the CMB with both the measurements of the temperature and of the polarization of the radiation, the Planck experiment has been able to yield two outcomes, essential for the establishment of the cosmological standard model, on the one hand, the original map of the CMB, not distorted by the lensing, that can be used as the input data in simulations, and, on the other a full sky map of the gravitational lensing potential which is tentatively interpreted in \cite{GCT1} and \cite{GCT2}  as the (anti-de Sitter) world matter identified with dark matter.  

\subsection{Simulation, a gedanken cosmological experiment algorithmically performed}

\begin{figure}[H]
\includegraphics[width=13.5 cm]{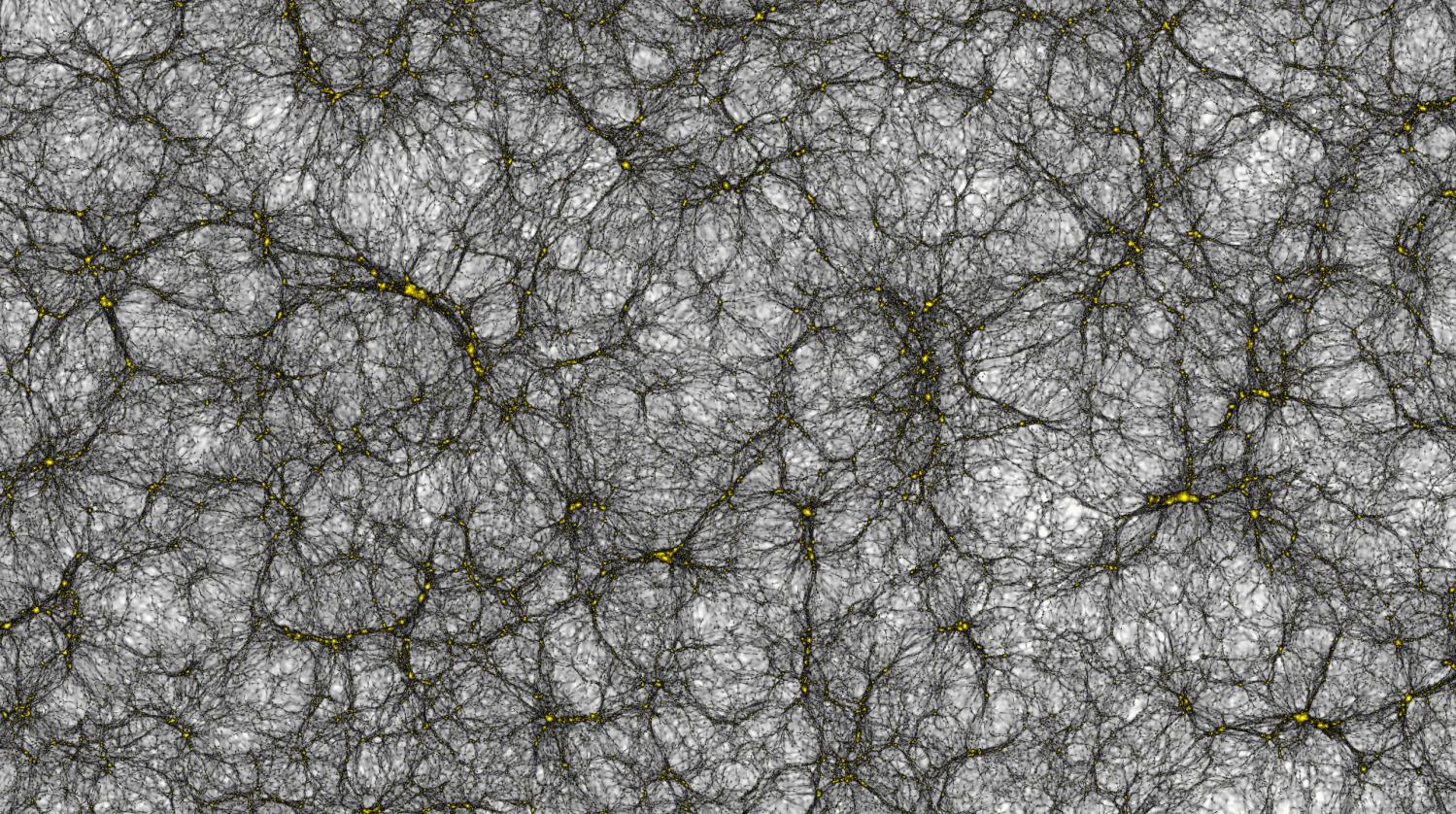}
\caption{A section of the largest virtual universe ever simulated. From \cite{physorg} \label{virtun}}
\end{figure}

The results of simulation, that can be considered as an algorithmic performance of the final stage of the cosmological gedanken experiment \cite{physorg}, are particularly spectacular. The figure \ref{virtun} can be interpreted as showing the complex topology of the spacetime of the dark universe: a web of dark filaments that are tensionless dark strings freely moving in a void space (the white regions in the figure) with negative curvature related to the cosmological constant, whereas the spacetime inside the filaments has a positive curvature.
	
\subsection{dS/AdS quantum elementary systems in Wigner's sense}
\label{WMdSAdS}
\subsubsection{dS and AdS geometries}
Here we abandon  the cosmological conception of $\Lambda$ as a part (pressure or density) of the right side  of the Einstein equation to instead adopt the fundamental constant point of view according to the place of $\Lambda$ should lie on the left of the Einstein equation, as it was discussed in   \cite{gaznov11}. 
Minkowski, de Sitter (dS)  and anti-de Sitter (AdS)  space-times are  maximally symmetric. 
dS and AdS  symmetries are  one-parameter deformations\\ \cite{bacryjmll68} of Minkowski symmetry. In terms of the cosmological constant $\Lambda$ we respectively have (in this dS/AdS context $c$ is restored)
 \begin{itemize}
  \item dS negative curvature  $-\sqrt{\Lambda_{\mathrm{dS}}/3}$\,,
  \item AdS positive curvature   $\sqrt{\vert\Lambda_{\mathrm{AdS}} \vert/3}$.
\end{itemize} 

The corresponding  kinematical groups are the 
proper orthochronous Poincar\'e group $\R^{1,3}\SD\, \mathrm{SO}_{\scriptsize 
0}(1,3)$ (or $\R^{1,3}\SD\, \mathrm{SL}(2,\C)$)
 the  dS ~SO$_{\scriptsize 0}(1,4)$ (or Sp$(2,2)$) and AdS ~SO$_{\scriptsize 0}(2,3)$ (or Sp$(4,\R)$) groups.

 dS space-time is conveniently represented by the
one-sheeted hyperboloid embedded in the $5$d Minkowski
space 
    $\mathsf{H}_{\mathrm{dS}} \equiv \{x \in \R^5 ;~x^2=\eta_{\alpha\beta}~ x^\alpha
x^\beta =-\Lambda_{\mathrm{dS}}/3\},   
    \, \ \alpha,\beta=0,1,2,3,4,
$ where $\eta_{\alpha\beta}=$diag$(1,-1,-1,-1,-1)$.

AdS  is represented by the 
one-sheeted hyperboloid embedded in  $\R^5$ equipped  with the metric:
    $\mathsf{H}_{\mathrm{AdS}} \equiv \{x \in \R^5 ;~x^2=\eta_{\alpha\beta}~ x^\alpha
x^\beta =\vert\Lambda_{\mathrm{AdS}}\vert/3\}\, ,   \
 \alpha,\beta=0,1,2,3,5,
$
where $\eta_{\alpha\beta}=$diag$(1,-1,-1,-1,1)$.

The Lie algebras of  groups dS and AdS group  are generated by the ten Killing vectors
$
\mathrm{K}_{\alpha \beta} = x_{\alpha}\partial_{\beta} -
x_{\beta}\partial_{\alpha}$. 

There exists a crucial difference between dS and AdS with regard to the question of time.   While there is no globally time-like Killing
vector in dS,   there is one in AdS, namely $K_{50}$. This fact has heavy consequences for attempting to properly define ``energy at rest'' in dS, as is shown below.

\subsubsection{Compared  classifications of Poincar\'e, dS and AdS  UIR's for quantum elementary systems}
\label{UIR}
In a given unitary irreducible representation (UIR) of  dS and AdS groups  their respective generators
map to  self-adjoint operators in Hilbert
spaces of spinor-tensor valued fields on dS and AdS respectively:
  \begin{equation*}
\mathrm{K}_{\al \be} \mapsto \mathrm{L}_{\al \be} =\mathrm{ M}_{\al \be} +\mathrm{ S}_{\al \be}\,,
  \end{equation*}
with orbital part  $\mathrm{M}_{\al \be}=-\ii (x_{\al}\partial_{\be} -
x_{\be}\partial_{\al})$ and spinorial part $\mathrm{S}_{\al \be}$ acting on
the field components. 

The physically relevant UIR's of the Poincar\'e, dS and AdS groups are denoted by $\mathcal{P}^{>}(m,s)$ (``$>$'' for positive energies), $U_{\mathrm{dS}}(\vsd,s)$, and $U_{\mathrm{AdS}}(\vsa,s)$, respectively. These UIR's are specified by
 the spectral values $\lg\cdot\rg$  of their quadratic and quartic  Casimir operators. The latter define two invariants, the most basic ones being predicted by the  relativity principle, namely proper mass $m$ for Poincar\'e and  $\vsd$, $\vsa$ for dS and AdS respectively, and spin $s$ for the three cases (details on their respective ranges are given in \cite{gazeau1-20}). 
 \begin{itemize}
\item	For Poincar\'e the Casimir operators  are fixed as
 \begin{equation}
\label{casP}
\begin{split}
\mathrm{Q}^{(1)}_{\text{Poincar\'e}} &= \mathrm{P}^{\mu}\,\mathrm{P}_{\mu} = {\mathrm{P}^0}^2 - \mathbf{P}^2= m^2\, c^2\, ,\\ \mathrm{Q}^{(2)}_{\text{Poincar\'e}} &= \mathrm{W}^{\mu}
\mathrm{W}_{\mu}  =  -m^2\, c^2\, s(s+1)\hbar^2, \ \mathrm{W}_{\mu} := \frac{1}{2}\epsilon_{\mu \nu \rho \sigma}
\mathrm{J}^{\nu \rho}\mathrm{P}^{\sigma}\,. 
\end{split}
\end{equation}
\item	 For de Sitter, 
\begin{equation}
\label{casdS}
\begin{split}
\mathrm{Q}_{\mathrm{dS}}^{(1)} &= - \frac{1}{2} \mathrm{L}_{\al \be}\mathrm{L}^{\al \be} = \vsd^2 -\left(s-\frac{1}{2}\right)^2 +2\equiv  \langle \mathrm{Q}_{\mathrm{dS}}^{(1)} \rangle\,, \\
\mathrm{Q}_{\mathrm{dS}}^{(2)} &= - \mathrm{W}_{\al}
\mathrm{W}^{\al}= \left(\vsd^2+ \frac{1}{4}\right)s(s+1) \,, \quad \mathrm{W}_{\al} := - \frac{1}{8}\epsilon_{\al \be \ga \delta \eta}
\mathrm{L}^{\be \ga}\mathrm{L}^{\delta \eta}\,.
\end{split}
\end{equation}
\item	For Anti-de Sitter, 
\begin{equation}
\label{casAdS}
\begin{split}
\mathrm{Q}_{\mathrm{AdS}}^{(1)} &= - \frac{1}{2} \mathrm{L}_{\al \be}\mathrm{L}^{\al \be}= \varsigma_{\mathrm{AdS}}(\varsigma_{\mathrm{AdS}} - 3) + s(s+1)\equiv \langle \mathrm{Q}_{\mathrm{AdS}}^{(1)} \rangle\, ,\\
\mathrm{Q}_{\mathrm{AdS}}^{(2)} &= - \mathrm{W}_{\al}
\mathrm{W}^{\al}= -(\varsigma_{\mathrm{AdS}}-1)(\varsigma_{\mathrm{AdS}} - 2)s(s+1\,, \ \mathrm{W}_{\al} := - \frac{1}{8}\epsilon_{\al \be \ga \delta \eta}
\mathrm{L}^{\be \ga}\mathrm{L}^{\delta \eta}\,. 
\end{split}
\end{equation}
\end{itemize}
While the relation between mass and energy in Minkowski is not ambiguous, these notions  in de Sitterian/Anti-de Sitterian geometry have to be devised from a flat-limit viewpoint, i.e. from the study of the contraction limit
$\Lambda\to 0$  of these representations. 
In this respect, a mass formula for dS has been established  by Garidi \cite{garidi03a}:
\begin{equation}
 \label{garidimass}
m^2_{\mathrm{dS}}:=\frac{\hbar^2\Lambda_{\mathrm{dS}}}{3c^2} (\langle Q_{\mathrm{dS}}^{(1)} \rangle - 2)=\frac{\hbar^2\Lambda_{\mathrm{dS}}}{3c^2} \left(\varsigma^2_{\mathrm{dS}}+ \left(s-\frac{1}{2}\right)^2\right)\,.
\end{equation}
This definition should be understood through  the contraction  limit of representations:
  $$
  \mathrm{dS}\, \mathrm{UIR}  \longrightarrow \mbox{Poincar\'e}\, \mathrm{UIR} \,. 
  $$ 
  More precisely, with
\begin{equation}
\label{dsnumass1}
\Lambda_{\mathrm{dS}}\to 0 \, \quad  \varsigma_{\mathrm{dS}}\to \infty\, , \quad \text{while fixing}\quad \varsigma_{\mathrm{dS}} \hbar \sqrt{\Lambda_{\mathrm{dS}}}/\sqrt{3}c= m_{\text{Poincar\'e}} \equiv m\, .
\end{equation}
we have 
\begin{equation}
\label{dscontr1}
U_{\mathrm{dS}}(\varsigma_{\mathrm{dS}},s)\underset{\vert \varsigma_{\mathrm{dS}}  \vert \sqrt{\Lambda_{\mathrm{dS}}}/\sqrt{3} = \frac{mc}{\hbar} }{\underset{\Lambda_{\mathrm{dS}}\to 0\,,\, \vert \varsigma_{\mathrm{dS}}  \vert\to \infty}{\lga}} {c_>\mathcal{P}}^{>}(m,s)
\oplus c_<\mathcal{P}^{<}(m,s)\,.
\end{equation}
This  result was proved in  \cite{micknied72} and discussed in \cite{garidi03}.
 One should  notice the possible breaking of dS irreducibility into a direct sum of
two Poincar\'e UIR's with positive and negative energy respectively. To some extent the choice of the  factors $c_<, \, c_>$, is left to a ``local tangent'' observer. In particular one of these factors can be fixed to 1 whilst the other one is forced to vanish. This crucial dS feature originates from  the dS  group symmetry 
mapping any point $(x^0, \mathsf{P}) \in \mathsf{H}_{\mathrm{dS}}$ 
 into its mirror image $(x^0, -\mathsf{P}) \in \mathsf{H}_{\mathrm{dS}}$ with respect
to the $x^0$-axis. 
Under such a symmetry the four dS generators
$\mathrm{L}_{a0}$, $a = 1,2,3,4$, (and particularly $\mathrm{L}_{40}$ which contracts
to energy operator!) transform into their respective opposite
$-\mathrm{L}_{a0}$, whereas the six $\mathrm{L}_{a b}$'s remain unchanged.

  Concerning AdS a  mass formula  similar to \eqref{garidimass}
 has been given  in\\  \cite{gaznov08,gaznov11}:
\begin{equation}
\label{adsgaridimass}
\begin{split}
m^2_{\mathrm{AdS}}&= \frac{\hbar^2 \vert\Lambda_{\mathrm{AdS}}\vert}{3c^2}\left(\langle Q^{(1)}_{\mathrm{AdS}}  \rangle- \langle Q^{(1)}_{\mathrm{AdS}} \vert_{\varsigma_{\mathrm{AdS}}=s + 1}\rangle\right)\\
& =  \frac{\hbar^2 \vert\Lambda_{\mathrm{AdS}}\vert}{3c^2}\left[\left(\vsi_{\mathrm{AdS}} - \frac{3}{2}\right)^2 - \left(s - \frac{1}{2}\right)^2\right] \,.
\end{split}
 \end{equation}
One here deals with the AdS group representations   $U_{\mathrm{AdS}}(\varsigma_{\mathrm{AdS}},s)$ with $\varsigma_{\mathrm{AdS}} \geq s+1$ (discrete series and its lowest limit), and their contraction limit holds with no ambiguity 
\begin{equation}
\label{adsnumass1}
 U_{\mathrm{AdS}}(\varsigma_{\mathrm{AdS}},s) \underset{\vsi_{\mathrm{AdS}}\sqrt{\vert\Lambda_{\mathrm{AdS}}\vert/3}= \frac{mc}{\hbar} }{\underset{\Lambda_{\mathrm{AdS}}\to 0\,, \,\varsigma_{\mathrm{AdS}} \to \infty}{\lga}} \mathcal{P}^{>}(m,s)\,. 
\end{equation}
Now, the  contraction formulae \eqref{dscontr1} and \eqref{adsnumass1} give us the freedom to write
\begin{equation}
\label{mmm}
m_{\mathrm{dS}}= m_{\mathrm{AdS}} = m\, , 
\end{equation}
which agrees with the Einstein position  that the proper mass of an elementary system should be independent of the geometry of space-time, or equivalently there should not exist any difference between inertial and gravitational mass.

 Let us now disclose a property of AdS which is essential for our interpretation of dark matter in our universe. Since the invariant $\zeta_{\mathrm{AdS}}$ is the lowest value of the discrete spectrum of the AdS  time generator we define the positive rest energy  as 
 \begin{equation}
E_{\mathrm{AdS}}^{\mathrm{rest}} := \hbar c\sqrt{\frac{\vert\Lambda_{\mathrm{AdS}}\vert}{3}}\,\zeta_{\mathrm{AdS}}\,.  
\label{defrestAdS}
\end{equation}
It results from Equation \eqref{adsgaridimass}:
\begin{equation}
\label{restenAdS}
E_{\mathrm{AdS}}^{\mathrm{rest}} = \left[m^2c^4 + \hbar^2 \omega^2_{\text{\scriptsize AdS}}\left(s - \frac{1}{2}\right)^2\right]^{1/2} + \frac{3}{2}\hbar \omega_{\text{\scriptsize AdS}}\, , 
\end{equation}
 with frequency $\omega_{\text{\scriptsize AdS}}:= \sqrt{\frac{\vert\Lambda_{\mathrm{AdS}}\vert}{3}}\,c$. 
Hence, to the order of $\hbar$, an AdS  elementary system in the Wigner sense is a deformation of
both a  relativistic free particle with rest energy $mc^2$ and a $3$d isotropic quantum harmonic oscillator with ground state energy $ \frac{3}{2}\hbar \omega_{\mathrm{AdS}}$ and with excited states which, apart from degeneracy, are spaced at equal energy intervals of $ \hbar \omega_{\mathrm{AdS}}$. 
 A complete proof of this feature in the $1+1$ AdS case is given in \cite{gare93}.

We do not find such a limpid result with dS. Nevertheless let us formally define 
 \begin{equation}
E_{\mathrm{dS}}^{\mathrm{rest}} := \hbar c\sqrt{\frac{\Lambda_{\mathrm{dS}}}{3}}\,\zeta_{\mathrm{dS}}\,,  
\label{defrestdS}
\end{equation}
which can assume any real value. The counterpart of \eqref{restenAdS} reads:
\begin{equation}
\label{restendS}
E_{\mathrm{dS}}^{\mathrm{rest}} = \pm\left[m^2c^4 - \hbar^2 c^2 \frac{\Lambda_{\mathrm{dS}}}{3}\left(s - \frac{1}{2}\right)^2\right]^{1/2} \, .  
\end{equation}
There is a noticeable simplification in both cases for spin $s = 1/2$ :
\begin{align}
\label{dsmdemi}
\mbox{for dS : }&\quad E_{\mathrm{dS}}^{\mathrm{rest}} =\pm mc^2\,, \\
 \label{adsmdemi}\mbox{for AdS: } & \quad  E_{\mathrm{AdS}}^{\mathrm{rest}} = mc^2 + \frac{3}{2}\hbar \omega_{\mathrm{AdS}}\,. 
\end{align}
The choice $E_{\mathrm{dS}}^{\mathrm{rest}} = mc^2$ should be privileged for obvious reasons. 
Moreover, in the massless case and spin $s$, we have
\begin{align}
\label{masslessdS}
\mbox{for dS : }&\quad E_{\mathrm{dS}}^{\mathrm{rest}} =\pm \ii \hbar \sqrt{\frac{\Lambda_{\mathrm{dS}}}{3}} c \left(s- \frac{1}{2}\right)\,,\\
\label{masslessAdS} \mbox{for AdS: } & \quad  E_{\mathrm{AdS}}^{\mathrm{rest}} = \hbar \sqrt{\frac{\vert\Lambda_{\mathrm{AdS}}\vert}{3}} c (s+1)\,. 
\end{align}
Therefore, while for dS  the energy at rest makes sense only for massless fermionic systems and is just zero, on the contrary, for AdS the energy at rest makes sense and is strictly positive  for any spin, and in particular for spin 1 massless bosons we get 
\begin{equation}
\label{AdSm0s1}
E_{\mathrm{AdS}}^{\mathrm{rest}} = 2\hbar \omega_{\mathrm{AdS}}\,. 
\end{equation}
One should add that in the  AdS massless cases with $s>0$ Gupta-Bleuler \& gauge structures have to be introduced in the description of quantum states. The spectrum of the time generator $L_{05}$ is still of the harmonic type, but there are  possible different degeneracies. Moreover the concept of helicity in AdS has to be reconsidered in terms of conformal symmetry. Details are given in  \cite{fronsdal75}  \cite{bifroheid83} \cite{gahamur89}. One important point to notice is that the quantum states in the massless cases  $\zeta_{\mathrm{AdS}} =s+1$, are described in holographic terms of vector-valued functions on the three-dimensional  Shilov boundary of the Cartan classical domain $\mathcal{R}_{IV}$ of the fourth type \cite{hua63} whereas the massive cases $\zeta_{\mathrm{AdS}} > s+1$ are described in terms of holomorphic functions in $\mathcal{R}_{IV}$.  
The latter is  diffeomorphic to the left group coset Sp$(4,\R)$/K, where K is the maximal compact subgroup  $\mathrm{S}(\mathrm{U}(1)\times \mathrm{SU}(2))$. Its Shilov boundary is diffeomorphic to $[0,\pi]\times\mathbb{S}^2$ (``Lie sphere''), or equivalently to the null cone in  $\R^5$ equipped with the $(+,-,-,-,+)$ metric. The latter can be viewed as the future horizon of the AdS space time.   

The existence of a strictly positive at rest energy for massless systems in AdS should not misinterpreted. For instance, the  energy in \eqref{AdSm0s1} might be viewed as a kind of ``mass gap''. However, the gluons, like all   AdS elementary systems for which $\zeta_{\mathrm{AdS}}= s+1$ and so $\langle \mathrm{Q}_{\mathrm{AdS}}^{(1)} \rangle= 2(s^2-1)$, are rigorously massless and propagate on the light cone, as proved for instance in the appendix of \cite{flafroga86}. Even though the ground state energy that they acquire in that AdS environment may be assimilated to that mass gap or/and energy gap, they remain conformally coupled, and do not experience any decay.

\subsubsection{From the elementary quantum context to the quantum cosmological context}

All what has been done in the present section in the elementary quantum context can be transposed in the cosmological context by coming back to the standard conception considering the cosmological term as a part of the right hand side of the Einstein’s equation. The phenomenological description of the matter content of the universe in terms of a perfect fluid characterized by a density and a pressure involves, in a four dimensional Minkowskian spacetime, a thermodynamical interpretation of the Friedman Lemaître differential equations that assimilates the boundaries in the far future (point $\omega$ in Figure \ref{inflatexp2} below, in which the Hubble radius     $L(a)={{H}^{-1}}(a)$ is plotted versus the scale factor   $a(t)$ in logarithmic scale) and the remote past (point $\alpha$ in Fig. \ref{inflatexp2}) implied by the Hubble expansion to \emph{event horizons with quantum properties} \cite{padmana03}. The future event horizon occurs in the region where the dark energy (attributed to CC) dominates, which leads us to call it a \emph{de Sitter horizon}; and, for reasons  that will appear clearer below we call the past event horizon, an \emph{anti-de Sitter horizon}. 

The methodology underlying this phenomenological description is the one of the \emph{effective theories} according to which, if there are parameters very large or very small with respect to the quantities of physical interest (with the same dimensions), one can  integrate out the very small and/or very large parameters and obtain a simpler, approximate description (said semi-classical) of the phenomena in terms of a family of effective theories depending only on finite but \emph{variable} effective parameters (said running or comoving): so, in our interpretation,   the de Sitter and anti-de Sitter world matter densities are meant to be \emph{effective co-moving world matter densities}.

\section{Matching the standard models of particle physics and cosmology}
\label{HPCOSMO}
\subsection{Our interpretation of the assets of $\Lambda$CDM}

\subsubsection{From time dependent densities to effective co-moving densities}
\begin{figure}[H]
\includegraphics[width=13 cm]{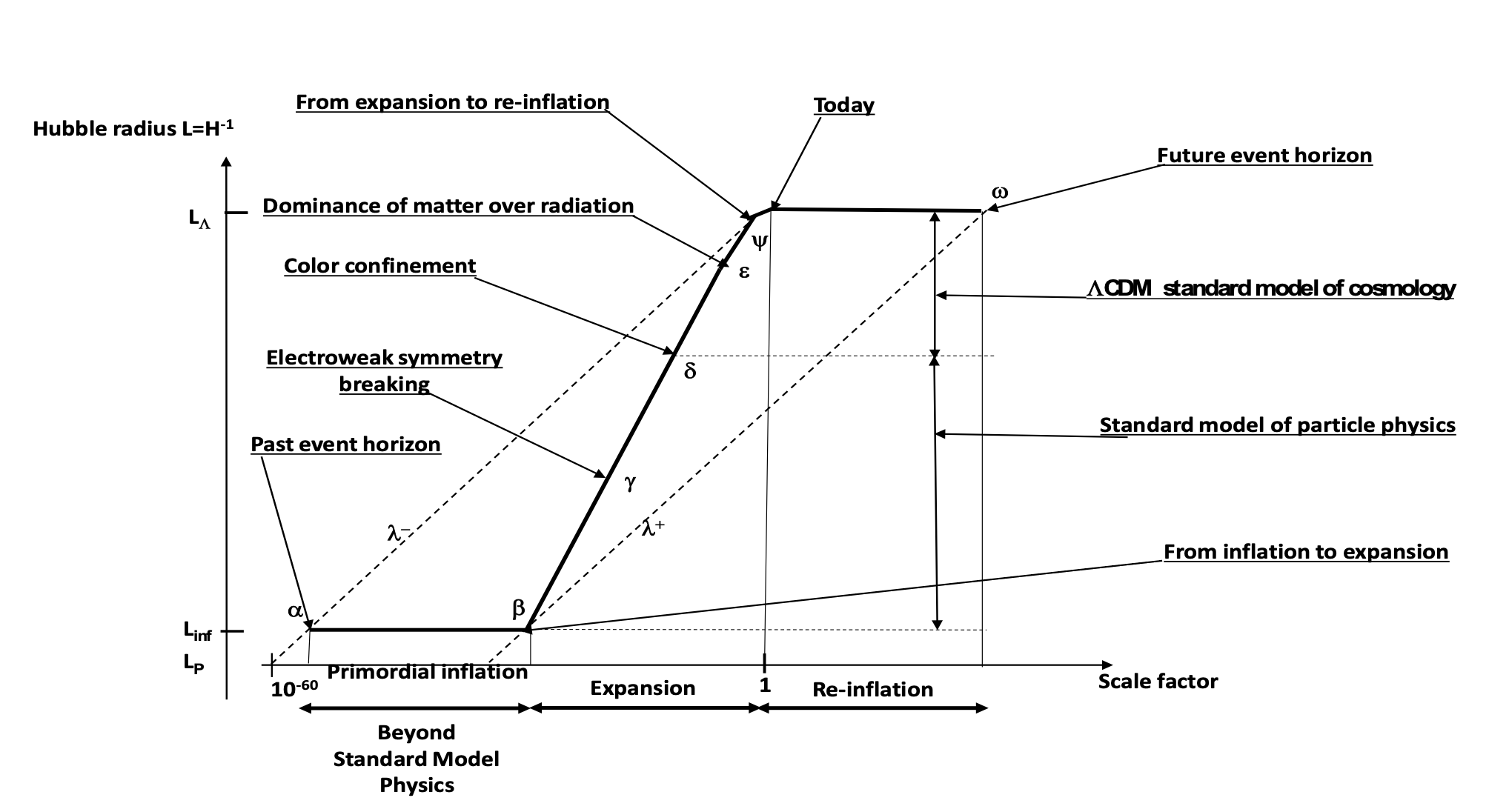}
\caption{Hubble radius     $L(a)={{H}^{-1}}(a)$ ($c=1$) is plotted versus the scale factor   $a(t)$ in logarithmic scale (from \cite{GCT1}).\label{inflatexp2}}
\end{figure}

The way how we interpret the assets of $\Lambda$CDM in terms the co-moving de Sitter and anti de Sitter world matter densities is illustrated  
by Figure \ref{inflatexp2}.

   The cosmic evolution is schematized on the thick line, on which the cosmic time, that is proportional to the logarithm of the scale factor, is made implicit, which allows somehow to solve the ``problem of time in cosmology'' by replacing all dimensioned quantities depending on the local time $t$,  by ``effective co-moving densities'' that are scaled by the scale factor depending on a global time, denoted $\tau$, which is said  \emph{thermal} \cite{coro94} because it depends through the Unruh’s constant   $U=\dfrac{h}{k_B c}$ on the temperature, at a power corresponding to their dimensions.
 
 In particular, this means that the dS and AdS curvatures discussed above in the quantum elementary context have to be replaced, in the quantum cosmological context by effective co-moving curvatures. For instance, the Hubble radius between $\beta$ and $\epsilon$  behaving like $a^2$ must be rescaled by a factor $a^{-1}$ because it is a length. The boundary of the Hubble $3$-sphere is a \emph{co-moving horizon}. So, \emph{the comoving Hubble radius behaves like the comoving radius of the universe}, which means, in terms of densities, that the number of degrees of freedom in the bulk equals the number of degrees of freedom on the boundary. It turns out that this ``holographic'' relation can be extended to the whole region between $\epsilon$  and  $\psi$  in which pressure-less matter dominates over radiation, in such a way that holography  is at work in the full expansion region from $\beta$  to  $\psi$.

 \subsubsection{Our interpretation of the flatness sum rule}
   \label{interflat}
   All quantum fluctuations exit from the co-moving  horizon in the primordial inflation phase, enter it in the expansion phase and re-exit it in the late inflation phase. No information-carrying quantum fluctuation with a wavelength smaller than   ${{\lambda }^{-}}$ on the left of the \emph{past informational}   $\alpha $ (this is the reason why we called it above an  anti-de Sitter horizon), or with  a wavelength larger than  ${{\lambda }^{+}}$  on the right of the \emph{future informational, or de Sitter  event horizon} $\omega $ enters the co-moving horizon. Note that the scenario of a  bouncing phase taking place at the point $\alpha$ has been proposed in recent works co-authored by one of us (see the review \cite{berczugamal20} and references therein).
   
   The interpretation of the holographic relation, that is at work in the full expansion phase from point $\beta$ to point $\psi$ is particularly clear at point $\psi$  that marks the transition from the expansion phase to the re-inflation phase at which the function $R(t)$    presents an inflexion point ($\ddot R=0$) which, through the second Friedman equation   \eqref{friedlem2} leads to equate the total bulk energy, or total active mass,  with the contribution of the cosmological constant (CC). It is clear, since the pressure associated with $\Lambda$ is negative, that this cannot be realized without a contribution with a positive pressure, that is with an anti-de Sitter world matter $\rho _{\text{AdS}}^{\text{ind}}(\psi )$ which exactly cancels at point $\psi$   the contribution of CC.
Such an anti-de Sitter world matter can be interpreted as the constant of integration resulting of integrating out the wave lengths smaller than ${{\lambda }^{-}}$, namely beyond the anti-de Sitter horizon.  When applied in the full expansion phase, the cancellation of CC by this  anti-de Sitter world matter amounts to replace in our quantum cosmology, the local time $t$ by the global time $\tau$. One could say that considering these two times amounts to a complexification of the time, and that $t$ and $\tau$ are complex conjugate variables: if the densities in our quantum cosmology are analytic functions depending on the global time $\tau$ they do not depend on its complex conjugate, namely the local time. We shall come back to this idea of the complexification of time in our contribution to the incoming book \cite{leshar21} on \emph{Time in Science}.

	More generally, the flatness sum rule that expresses the vanishing of the spatial curvature \cite{GCT2} equates the sum of the visible energy  density $\rho _{\text{vis}}$ (the baryonic $\rho _{\text{b}}$  and radiative $\rho _{\text{R}}$ energy densities), the dark energy density and the dark matter energy density, which amounts to nothing but the total active mass in the effective comoving dark universe with a radius equal to its Hubble radius,  to the so called critical density $\rho _{\text{c}}= \dfrac{3H^2}{8\pi G_{\text{N}}}$. The latter  is the energy density at the boundaries in the far past and in the far future of the Hubble horizon in the absence of any integration constant and any spatial curvature. More precisely, this flatness sum rule reads as:
\begin{equation}
\label{balrho1}
\rho _{\text{vis}}+ \rho _{\mathrm{DM}}+\rho _{\text{DE}} -\rho _{\text{c}}=0\,, 
\end{equation}
with
\begin{equation}
\label{balrho2}
 \rho _{\text{vis}}= \rho _{\text{b}}+\rho _{\text{R}}\, , \quad  \rho _{\text{DE}}=\frac{\Lambda}{8\pi G_N}\,. 
\end{equation}	 
Our interpretation 
is inspired by the seminal work of Brout, Englert and Gunzig \cite{broutetal78} which states 
\begin{quote}
\textit{Cosmology, because it is concerned with the variation of  $g_{\mu\nu}$ within a distribution of matter and not without, is described - at least in the mean - by only that part of $g_{\mu\nu}$ which is its determinant that may be represented by a scalar field $\phi$ in Minkowski space .}
\end{quote}
This leads us to equate in Eq. \eqref{balrho1} the critical density $\rho _{\text{c}}$ to the energy density $\rho _{\phi}$ of the so-called \textit{dilaton}, i.e., the covariant (comoving) quantum field $\phi$, representing the determinant of the Friedman, Robertson, Walker (FRW) metric of the effective comoving dark universe.  
According to our methodology of effective field theory, $\phi$ has the equation of state $W_{\phi}= P_{\phi}/\rho_{\phi}= -1/3$. Thus $\rho _{\text{c}}= \rho_{\phi}= -3P_{\phi}$ and this  insures the vanishing of the total active mass of the vacuum, the zero point of energy: with
\begin{equation}
\label{balrho3}
 \rho _{\text{vis}}+ \rho _{\mathrm{DM}}\equiv \rho _{\mathrm{M}}= \frac{1}{2} \rho _{\text{DE}}=-P_{\phi}\, , 
\end{equation}
Eq. \eqref{balrho1} becomes, with all densities and pressures (including $\rho _{\text{DE}}$) being rescaled at time $\tau$,
\begin{equation}
\label{relinert}
 \rho _{\text{vis}}+ \rho _{\mathrm{DM}}  +  \rho _{\mathrm{DE}} = -3P_{\phi} = \rho _{\phi}\,.
\end{equation}
This means that the dark matter density and the dark energy sum to what must be added to the visible matter density in order, for the total matter-energy density, to satisfy the principle of the relativity of inertia $\rho _{\mathrm{M}}+ P _{\phi} =0$, or the vanishing of the spatial curvature $ \rho_{\phi}+3P_{\phi} = 0$.

An important remark is in order here: it may look puzzling that, whereas the equations of states (EoS) of both dark matter and dark energy are equal to $-1$, the EoS of the effective comoving dark universe is equal to $-1/3$.  But, in \cite{melia17} F. Melia has solved this puzzle: the EoS of energy densities (that are not true scalar energy densities) are frame dependent; when the active mass does not vanish, a comoving frame does not, in general, coincide with a free fall frame, (that may happen only at one particular position), but ``in FRW, the comoving and free-falling frames are supposed to be identical at every spacetime point.'' 

 \subsubsection{The primordial inflation and the minimal ``beyond the standard model'' (BSM) assumption}
 
 In $\Lambda$CDM, the primordial inflation phase occurs at a Hubble radius of about $10^3$ to $10^4$ Planck’s lengths, which is clearly a domain of physics beyond the standard model (BSM) and supposedly relying on quantum gravitational effects. It is generally admitted that, although it does not resolve completely the cosmological singularity problem (the singularity is sent behind an event horizon) the primordial inflation phase helps correcting the well-known defects of the ``simple big bang model'', namely the absence of monopoles, the vanishing of the spatial curvature and the particle-horizon problem. It ends with the emission of particles with a mass of about $2.5$ meV, which could be neutrinos, and which, as is frequently assumed,  we associate with the cosmological constant and thus to the dark energy. The scale at which it occurs has long been associated with the grand unification symmetry breaking or supersymmetry breaking, but the fact that it corresponds to the zero point of the energy strongly suggests that it has rather to be associated with the matter/antimatter symmetry breaking: the right hand side  of Eq. \eqref{balrho3}, half of the dark energy density, is nothing but the missing dark energy, the one of the anti-matter that has been ``integrated out''. We remind that if matter would reduce to visible matter (which is supposed to be pressure-less) there would be no negative pressure  to insure the principle of the relativity of inertia. It is the addition of the dark matter density that may insure the validity of this principle. We shall see below that this feature is an essential key point of our paper.

To link the scales of the primordial inflation, of the cosmological constant, of the neutrino masses and of the matter/antimatter symmetry breaking is indeed very appealing. To explore the possible consequences of such an association, one must discuss the problem of the neutrino masses per the Brout Englert Higgs (BEH) mechanism. It is well known that the standard model using the BEH mechanism is compatible with massless neutrinos. In fact, right-handed neutrinos which would be necessary to generate mass through the Yukawa coupling of the BEH boson to the right-handed and left-handed neutrinos, have quantum numbers which make of them \emph{SM sterile} particles (zero weak isospin, zero charge and zero weak hypercharge). So, the SM is completely compatible with massless neutrinos. If neutrinos are massive, as they seem to be, their mass is thus highly likely a signal of BSM physics. The simplest assumption is to assume that there do exist sterile right-handed neutrinos. The problem is that if one gives the neutrinos Dirac masses through the Yukawa couplings to the BEH boson, one does not understand why these masses are much smaller than the Dirac masses of the charged fermions. The way out of this difficulty comes from the fact that the right-handed neutrinos can have a Majorana mass by their own. If the Majorana mass is exceptionally large, one can manage, with a ``seesaw'' mechanism to get, by diagonalizing the mass matrix, a physical neutrino which would be a linear combination of the normal left-handed neutrino plus a small anti-right-handed neutrino component. With a Majorana mass of the order of the primordial inflation scale, one can have neutrino masses in the milli-eV range in possible agreement with the results of neutrino oscillations experiments. Furthermore, this mechanism has the advantage that through their Yukawa couplings, the sterile righthanded neutrinos can decay into standard model particles (a BEH boson plus a lepton) thus providing a mechanism of lepton number non-conservation (leptogenesis) (see for instance \cite{ramond05,bupeya05}).  Now, through the so-called “sphaleron” mechanism, the breaking of lepton number can lead to the breaking of baryon number (baryogenesis) at the primordial inflation scale, thus satisfying one of the Sakharov’s conditions for the origin of matter \cite{sakharov67B}. We thus adopt this assumption which can be considered as the minimal BSM assumption about the initial condition of quantum cosmology.

\subsection{The dark matter induced by QCD}

\begin{figure}[H]
\begin{center}
\includegraphics[width=4in]{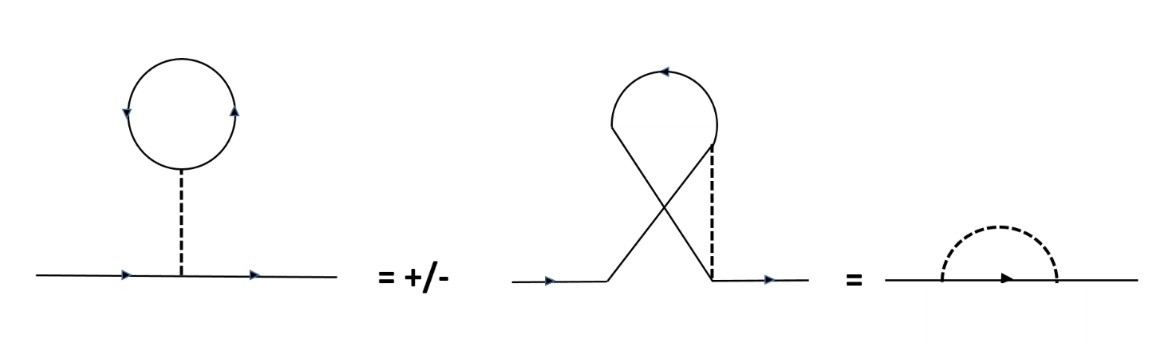}
\caption{A ``tadpole'' diagram in which a boson (resp. fermion) exchanges a virtual dilaton with a vacuum loop involving a particle identical to it, is transformed trough the interchange of identical particles, into a positive, i.e. increasing the mass (resp. negative, i.e. increasing the energy) self-energy diagram. \label{taddiag}}
\end{center}
\end{figure}

If one wants to match the standard models of cosmology and particle physics, one has to move  on the thick line of Figure \ref{inflatexp2}, either ``bottom up'', $\psi $ to point  $\delta$ that marks the transition from the QCD quark gluon plasma to the colorless hadronic phase, or, ``top down'' from point $\beta$ and through point $\gamma$, the electroweak symmetry breaking per the BEH mechanism, to  point $\delta$ that represents the low energy frontiers of the standard model of particle physics and the high energy frontiers of the one of quantum cosmology. Our idea is thus to interpret $\rho _{\text{AdS}}^{\text{ind}}$  as a comoving density that, when evaluated at point $\delta$, would plays the role of the anti-desitterian world matter, induced by QCD, to cancel the contribution of the comoving CC at point $\delta$. Now, it turns out that following an idea of Sakharov \cite{sakharov67} and the work of Adler \cite{adler82} such a contribution can be rigorously evaluated (or at least estimated) \cite{brindejonc95,brindejonc98}. The idea of Sakharov was that the non-renormalizable Einstein-Hilbert action would be an effective theory resulting from the coupling of a renormalizable gauge theory to a renormalizable gravitational theory quadratic in the curvature. The aim of Adler was to use the methodology of effective theories to evaluate the cosmological term  induced by integrating out, in the effective action, the quantum fields of the standard model:
\begin{equation}
\label{LGT}
-\frac{1}{2\pi}\frac{\Lambda_{\mathrm{ind}}}{G_{\mathrm{ind}}}= \frac{\int\ud \{\Phi\}\,e^{\ii S\left[\{\Phi\},\eta_{\mu \nu}\right]}T(0)}{\int\ud \{\Phi\}\,e^{\ii S\left[\{\Phi\},\eta_{\mu \nu}\right]}}\,, 
\end{equation}
where $T(0)$, the trace anomaly, can be evaluated in terms of the flat space time vacuum expectations of renormalized products of gauge and matter fields (called condensates). In QCD, 
these condensates involve a mass scale parameter $M(g,\mu)= \mu\exp(-1/b_0 g^2)$ where $\mu$ is the renormalization scale and $b_0=\left(11 N_c-2 N_f\right)$  where $N_c$ is the number of colors and $N_f$ the number of quark flavors, that plays, in QCD, the same role as the scale parameter $a(\theta)$  where $\theta$ is the temperature. The mass scale parameter presents an essential singularity at  $g^2=0$, so the induced cosmological term cannot be evaluated perturbatively. Anyhow, if one can use some non-perturbative technique such as the lattice gauge quantization, one can expect all the condensates contributing to the trace anomaly to be proportional, with a negative factor, to the constant $b_0$. For instance, the contribution of the  di-gluons through what is called in \cite{adler82} the \textit{gluon pairing amplitude}   to the trace anomaly reads
\begin{equation}
\label{Tmumu}
\left\lg T^{\mu}_{\ \, \mu}\right\rg_0= -\frac{1}{8}\left[11 N_c-2 N_f\right]\,\left\lg \frac{\alpha_s}{\pi}\left(F^a_{\mu\nu}F^{a\mu\nu}\right)^r\right\rg_0\,. 
\end{equation}
In the following quote from \cite{GCT2} it was argued: 
\begin{quote}
\emph{The minus sign in the right hand side shows that when the constant $b_0=\left(11 N_c-2 N_f\right)$  is positive,  all the QCD condensates contribute negatively to the energy density, which means that the QCD world-matter is globally an anti-de Sitter world-matter (dominance of an anti-de Sitter world-matter over a smaller de Sitter world-matter).} 

\emph{The multiplicative factor $b_0$ allows reading, thanks to the well-known property that boson and fermion loops contribute in quantum field theory with opposite signs, see Figure \ref{taddiag},  the relative contributions of the components of the QCD vacuum to the full world-matter:
\begin{itemize}
  \item the bosonic (gluon) loops, proportional to $N_c$, contribute to the anti-de Sitter world matter that represents the contribution of the bulk of free gluons to the total active mass in the effective dark universe, and 
  \item the fermionic (quark) loops, proportional to $N_f$, contribute to the normal de Sitter world matter, which, per our interpretation, represents the kinetic energy density of the quarks which  have survived to the global annihilation of fermions and antifermions, namely the constituents of the baryonic matter.
\end{itemize}}
\end{quote}
In this quotation of a previous work appears the expression of free gluons that we will not retain in the present paper because it is misleading: the gluons in the quark gluon plasma that can be considered as free before the hadronization transition, have to be ``integrated out'' (see Eq. \eqref{LGT}) to form, in the colorless hadronic phase, the non-interacting world matter induced by QCD that will be interpreted below (\ref{BECCC}) as  Bose-Einstein condensate of di-gluons.

Now, since the transition from the Quark Gluon Plasma (QGP) to the colorless hadronic phase occurs in the region of expansion in which we use the methodology of effective theories, we assimilate the full content of the universe at point $\delta$ to an ``effective dark universe'' for which the radius of the universe is equal to the Hubble radius. This means that we have (by thought) sent the baryonic matter at the Hubble horizon namely made of its energy density a de Sitter world matter. This is the key point of \cite{GCT2}:  the term in Eq. \eqref{Tmumu} proportional to $N_f$ is a de Sitter world matter that represents, at point $\delta$, the kinetic energy of the quarks, called ``valence quarks'' that constitute the baryonic matter, whereas the term proportional to $N_c$ is an anti de Sitter worldmatter that represents the active mass of the gluons, namely the dark matter. At point $\delta$, $N_c$ is equal to 3, and $N_f$, which is not the number of quark flavors, but rather the number of fermions that constitute a nucleon,  is also equal to 3. Eq. \eqref{Tmumu} thus allows us to conjecture the value of the ratio Dark/Visible to be equal to 11/2, at point $\delta$ but also today since pure numbers need not to be rescaled. This is the main outcome of \cite{GCT2} that we now confront with the results of the Planck experiment: in Table 2 of \cite{planck18VI}  these results are compared with the expectations from the so-called ``base LambdaCDM'' standard model. In the last column of this table, we can read the 68\% limits on the parameters based on “Planck TT,TE,EE,+lowE+lensing”: for $\Omega_{\Lambda}$ we read  $0.6889\pm 0.0056$,  for   $\Omega_{\mathrm{M}}$ (that stands for $\Omega_{\mathrm{vis}} + \Omega_{\mathrm{DM}}$)  we read  $0.3111\pm 0.0056$. Now, according to our model, we expect 
\begin{equation}
\label{OmOm}
    \Omega_{\Lambda}- \Omega_{\mathrm{vis}} =  \Omega_{\Lambda}-\frac{2}{11} \Omega_{\mathrm{M}} =2 \Omega_{\mathrm{M}}
         \Rightarrow \Omega_{\mathrm{M}} =  \frac{11}{24} \Omega_{\Lambda}\approx 0.3157\, , 
\end{equation}
to be favourably compared with $0.3111\pm 0.0056$.

 \subsubsection{Baryons as “chromo-magnetic” monopoles}

A superconductor analogy was used in \cite{nielsen79} by Nielsen and Olesen who proposed a suggestive model of the QCD vacuum involving unconfined chromo-magnetic monopoles moving freely along magnetic flux lines.  The interpretation of baryons as color magnetic monopoles\footnote{As a special tribute to Georges Lochak (1930-2021), French physicist known for his work on magnetic monopoles.} had been proposed by Ed. Witten \cite{witten79} in the following quote (where $N$ has to be replaced by $N_c$):
\begin{quote}
\emph{Indeed, the baryon mass is of order $N$, which can be written as $1/(1/N)$. But $1/N$ is the ``coupling constant'' of the strong interactions, which characterizes the interaction among mesons. $1/N$ plays in QCD roughly the role that $\alpha$ plays in spontaneously broken gauge theories of the weak and electromagnetic interactions. The fact that the baryon mass is of order $1/(1/N)$ is analogous to the fact that the Polyakov-'t Hooft monopole mass is of order $1/\alpha$.}
\end{quote}
.

 \subsubsection{Magnetic flux lines as dark matter filaments}
	
	In \cite{nielsen79} Nielsen and Olesen   have argued that ``one gains energy by separating a monopole and an anti-monopole''. It is thus reasonable to interpret the color magnetic flux lines (a pure QCD effect) as filaments of a world matter (that is a component of the universe with no interaction other than gravitational) that is filaments of dark matter connecting monopoles to anti-monopoles.  Now, since we can assume that anti-monopoles have been integrated out, in giant black holes at the center of galaxies or galaxy clusters, we expect that in simulations of the distribution of galaxies, the filaments must close at the points where are located the heavy black holes. And this is precisely what is seen in Figure \ref{virtun}.

\subsection{Bose Einstein condensation in the cosmological context}
\label{BECCC}
Since 1999 there is experimental evidence (RHIC, LHC) of the quark-gluon plasma as a super-liquid which is produced in ultra-relativistic heavy ion collisions, see for instance \cite{passum17,stachel18} and the recent comprehensive historical account \cite{rafelski20}. Measurements indicate that quarks, antiquarks and gluons flow independently in this liquid.
 On the other hand   the universe at its quark epoch, i.e. from $10^{-12}$s to $10^{-6}$s, with temperature $T > 10^{12}$K, was uniformly filled with QGP which once the Universe cooled below evaporated into a gas of hadrons. This corresponds to the point $\delta$ in Figure  \ref{inflatexp2}. As explained  above, an effective AdS  with curvature provided by  $\Lambda$ is present at this period. 

Returning to  our approach to elementary systems in dS or AdS space-times,  Equation \eqref{adsmdemi} tells us that the energy at rest of a fermion in an AdS background decomposes into a ``visible'' mass part, like in Minkowski, and a ``dark'' part which is like the ground state energy of a quantum three-dimensional isotropic harmonic oscillator with frequency equal to  $\sqrt{\dfrac{\left| {{\Lambda }_{\text{AdS}}} \right|}{3}}c$. This feature led one of us in  \cite{gazeau1-20} to infer that at the point $\delta$ , i.e., at the hadronization phase transition, ``chemical freeze-out'' temperature $T_{cf} \gtrsim 1.8\times10^{12}$K,  the ratio ``dark/visible'' $r:= \dfrac{3}{2}\dfrac{\hbar \omega_{\mathrm{AdS}}}{mc^2}$ for light quarks $u$ (mass $m_u\approx 2.2$\,MeV/$c^2$)  and $d$ ($m_d\approx 4.7$\,MeV/$c^2$) are given by $r(u)\approx 108$ and $r(d)\approx 49$ respectively.   In Reference \cite{gazeau1-20} it was suggested that dark matter originated from this supplementary mass granted to (anti-)quarks  by the ADS environment. In the present paper we instead explain the existence of dark matter as holding its origin from the gluonic component of the QGP.   On the same footing, we tentatively explain dark matter by asking a simple question:
what becomes the huge amount of gluons after the transition from QGP period to hadronization?

Equation \eqref{AdSm0s1} tells us that the energy at rest of a spin 1 massless boson in an AdS background is purely ``dark'' and is twice the elementary quantum $\hbar \omega_{\mathrm{AdS}}$. Hence the QGP gluons in the AdS background   at the point $\delta$ acquire an effective mass $2\hbar \omega_{\mathrm{AdS}}$. The latter is qualitatively determined through the equipartition  $k_BT_{cf}\approx \hbar \omega_{\mathrm{AdS}}$. Hence,   $2\hbar \omega_{\mathrm{AdS}}/c^2=144\times m_u\approx 317$\,MeV/$c^2$. One should notice that this gluonic effective mass is about $4/3$ times the effective mass acquired by quarks and antiquarks in that QGP-AdS environment. 

Now, it is tempting to establish a parallel between dark matter and  CMB, since the latter is viewed as the emergence of the photon decoupling, precisely when photons started to travel freely through space rather than constantly being scattered by electrons and protons in plasma. Hence, one may assert that a (considerable) part of the gluonic component of the quark epoch freely subsists after hadronization, within  an effective AdS environment,  as an assembly of a large number, say $N_G$,  of non-interacting entities that are  not individual free gluons, but rather decoupled gluonic colorless systems,  which are assumed to form a grand canonical Bose-Einstein ensemble. As said above, in \ref{interflat}, the holographic relation that holds in the expansion phase, from point $\beta$ to point $\psi$ in Figure \ref{inflatexp2}, cannot be satisfied without an anti-de Sitter world matter density that can compensate (see Eq. \eqref{Tmumu}) the contribution of CC in Eq. \eqref{friedlem2}. The simplest purely
gluonic system of this type, susceptible to form a Bose-Einstein condensate is a di-gluon whose contribution is the one of the gluon pairing amplitudes of Eq. \eqref{Tmumu}. The di-gluons can be called ``dark matter quasi-particles'', which, with an AdS rest mass equal to $\sqrt{\Lambda}$ form a Bose-Einstein condensate because their Compton wavelength  as well as their mean relative distances equal with the Hubble radius. Let us remind the reader that, according to the methodology of effective theories, all dimensioned quantities such as the rest mass of a quasi-particle are rescaled, i.e. depend on the global thermal time $\tau$ (see above \ref{interflat}).

Actually, the ``effective dark universe'' which, as said above, is supposed to provide the cosmological standard model with a  quantum vacuum or a ground state, cannot be thought of as an ``empty spacetime'' in which some \underline{objects} would move, but rather as a medium in which occur vacuum polarization \underline{events},  what G\"ursey \cite{gursey63} calls scintillation events\footnote{The mass scintillation model imagined by Gürsey, is comparable with the steady state cosmology of Bondi \cite{bondi48} in which the creation, at constant density of matter- energy, induces the expansion of the universe.}, namely events each consisting in  the virtual creation of a particle-antiparticle pair, followed, a short time later, by its annihilation. If the particles of the pair are fermions, such event would contribute to a negative curvature (normal de Sitter) world matter, a sort of a \emph{baryonic Fermi-Dirac sea}, at the exterior of the Hubble horizon, and if they are bosons, the event would contribute to a positive curvature (anti-de Sitter) world matter, a \emph{gluonic Bose-Einstein condensate}, at the interior of the Hubble horizon.

Therefore we consider an assembly of these $N_{G}$ dark matter di-gluons viewed as quasi-particles in an AdS environment  with individual energies $E_n= E_{\mathrm{AdS}}^{\mathrm{rest}} +  n\hbar\omega_{\mathrm{AdS}}$ where $E_{\mathrm{AdS}}^{\mathrm{rest}}$ is the expression \eqref{restenAdS} taken at $m=m_G$ (can be zero or negligible) and $s=0$, and degeneracy $g_n= (n+1)(n+3)/2$ \cite{fronsdal75}. Consequently,  those remnant entities are analogous to isotropic harmonic oscillators in $3$-space. They are assumed to form a grand canonical Bose-Einstein ensemble whose the chemical potential $\mu$ is fixed by the requirement that the sum over all occupation probabilities at temperature $T$ yields \cite{grossholt95,mullin97}
\begin{equation}
\label{NGBEC}
N_G= \sum_{n=0}^{\infty} \frac{g_n}{\exp\left[\frac{\hbar\omega_{\mathrm{AdS}}}{k_B T}\left(n+\nu_0 -\mu\right)\right] -1}\, , \quad \nu_0:= \frac{E_{\mathrm{AdS}}^{\mathrm{rest}}}{\hbar\omega_{\mathrm{AdS}}} \,.
\end{equation}
Since this number is very large one expects that this Bose-Einstein gas condensates  at temperature 
 \begin{equation}
\label{TBEC}
T_c \approx \frac{\hbar\omega_{\mathrm{AdS}}}{k_B} \left(\frac{N_G}{\zeta(3)}\right)^{1/3}\approx 1.18\times 10^{-3}\times \sqrt{\vert\Lambda_{\mathrm{AdS}}\vert}\,N_G^{1/3}
\end{equation} 
 to become the currently observed dark matter. The above  formula involving the value $\zeta(3)\approx1.2$ of the Riemann function is standard for all isotropic harmonic traps (see for instance \cite{grossholt95}, however, one should not be misled by the latter term: there is no harmonic trap here, it is the AdS geometry which originates the harmonic spectrum on the quantum level).     To support this scenario, and as explained in \cite{CCT05}, it has been shown, thanks to the physics of ultra-cold\footnote{As a side remark, one could propose to name the cosmological standard model $\Lambda$UCDM (UCDM for Ultra Cold Dark Matter)} atoms, that Bose Einstein condensation can occur in non-condensed matter but also in gas, and that this phenomenon is not linked to \emph{interactions} but rather to the \emph{correlations implied by quantum statistics.} 

Although we do not  precisely know at which stage beyond the point $\delta$ in Figure \ref{inflatexp2} does take place the gluonic Bose Einstein condensation,  let us  see if Eq. \eqref{TBEC} yields reasonable orders of magnitude by putting  $T_c$ equal to the current CMB temperature, $T_c= 2.78$K, and $\vert\Lambda_{\mathrm{AdS}}\vert \approx \frac{5.5}{6.5}\times \frac{11}{24}\times\Lambda_{\mathrm{dS}}= 0.39\times \Lambda_{\mathrm{dS}}$, with $\Lambda_{\mathrm{dS}}\equiv$ present CC $\Lambda= 1.1\times10^{-52}$m$^{-2}$.  We then derive from \eqref{TBEC} the following  estimate on the number of di-gluons in the condensate:
\begin{equation}
\label{NGcurrent}
N_G\approx 5\times 10^{88}\,. 
\end{equation} 
This estimate already seems  reasonable since the gluons  are around $10^{9}$ times the number of baryons, and the latter is estimated to be around $10^{80}$. 
Keeping this number in Eq. \eqref{TBEC} allows to estimate the scaling factor defined by
\begin{equation}
\label{scaleTL}
\sigma_c:= \frac{\Lambda_{\mathrm{dS}}}{T^2_c}\approx 1.85\times10^6\times N_G^{-2/3}\approx 1.36\times10^{-53}\,. 
\end{equation}
Hence, having  $T_c$ of the order of  ``Matter-dominated era''  temperature, say $10^4$K, i.e., at the point $\varepsilon$ in Fig. \ref{inflatexp2},  yields the following value for $\Lambda_{\mathrm{dS}}$: 
\begin{equation}
\label{}
\Lambda_{\mathrm{dS}}(\varepsilon)\approx1.36\times 10^{-44}\,\mathrm{m}^{-2}\,. 
\end{equation}

Some years ago,  it has been shown that Bose-Einstein condensation of dark matter can solve the core/cusp problem in the rotation curves of dwarf galaxies \cite{bohar07,harko11}  with dark matter particles of mass in the meV range \footnote{We would like to acknowledge the anonymous referee who let us know these references.}  In these motivating  articles the exact nature of the BEC particles is not made precise, but strong interaction between them is assumed, and 
dark matter is viewed as a non-relativistic, Newtonian Bose-Einstein gravitational condensate gas, whose density and pressure are related by a barotropic equation of state. Some  observed properties of dwarf galaxies appear to fit well with this scenario.   More recently, the interpretation of dark matter in terms of a Bose-Einstein condensate has drown interest in the framework of the so called ``fuzzy dark matter'' (FDM) model \cite{hbg00}. Originated from the idea that the dark matter particle is an ultralight particle \cite{hotw17}, the \emph{axion}, this model intends to solve the small scale shortcomings of the cold dark matter (CDM) \cite{dufbib09} model (``cuspy'' dark matter halo profiles and an abundance of low mass halos). In the FDM model, the mass of the would be dark matter particle must of the order of $10^{-22}$\,eV in order for its wave nature to be significant at astrophysical scales. For testing the ability of such a model to account for observations, it was necessary to use simulations.  Such high precision simulations were done in \cite{scb14} and they confirm the expected order of magnitude of $10^{-22}$\,eV for the mass of the dark matter particle. Whereas such small masses seem highly unrealistic for particles to be considered in particle physics, we think that our scalar covariant quantum ``dilaton'' field $\phi$, which is not a ``dark matter particle'', but rather, using a terminology of condensed matter physics,  a “\textit{dark matter quasi-particle, or a collective excitation}”   viewed as a di-gluon, can have a temperature-dependent mass gap of this order of magnitude. 

\section{Discussion}
\label{disc}

In this concluding discussion section, we would like, on the one hand, to see in which way the present paper could fill the gaps between our two former approaches and, on the other hand, to propose new arguments in favor of their common aim, namely the aim of reconciling the standard models of particle physics and cosmology. The starting point of the approach of GC-T was, as explained above, to associate dark matter with a gravitational lensing potential and thus, to assimilate it to an anti-de Sitter world matter. This approach was lacking a firmer theoretical ground, which, we think has been provided by the approach of J-PG. The idea of this approach is that, apart from the ordinary mass, elementary quantum systems can acquire in an anti-de Sitter space time a supplement of mass, of purely geometric origin, that could be a candidate, in the cosmological context, that is, as explained in subsection \ref{DMADSWM}, when the cosmological term in the Einstein’s equation is put back in the right hand side in terms of world matter densities, to constitute the dark matter. 
	
	Actually, the link between the two approaches can be provided by what we call the ``\emph{dilaton}'' field $\phi$, that is a scalar \emph{covariant} quantum field  which is given by \emph{non-perturbative} effects of the gauge theories of the standard model, at each cosmic event of emergence of masses, a ``Higgs like'' potential (with a Mexican hat shape) that can be interpreted as a world-matter energy density. This is indeed true at the electroweak symmetry breaking at point $\gamma$ in figure \ref{inflatexp2}, but also at the QCD hadronization transition at  point $\delta$, in Figure \ref{inflatexp2}. Let us note that, in \cite{dufbib09}, the authors attribute to the \emph{axion} the potential provided by the non-perturbative QCD effects, whereas we attribute it to the \emph{dilaton field}, which, as explained above looks much more reasonable. 
	
In particle physics, the frontier of the standard model is essentially due to non-perturbative effects such as instantons or monopoles that can be encountered in the cosmological context \cite{weinberg96}. Fortunately, as explained in \ref{interflat},  it turns out that cosmology at extra galactic scales may be, because it depends only on the determinant of the metric, easier  to deal with than at smaller scales.    The question to know whether the filament structure that appears in the distribution of dark matter at extra galactic is also present at smaller scales, is still an open question. For instance, it is often claimed that, at galactic scales, dark matter is rather distributed in the halo of ordinary matter. But we think that this feature cannot be objected as an argument against our approach, because our interpretation of dark matter as a Bose-Einstein condensate, precisely aimed to correct the defects of the conventional cold dark matter model, has been some how comforted, as shown above, by the fuzzy dark matter paradigm. On the other hand, it seems that a very recent work, based on deep learning, has shown the existence of a local web of dark matter at galactic scales \cite{hongetal21}.

\subsection*{Acknowledgments}
The authors thank Dr. Vincent  Brindejonc (Thal\`es) and Prof.  Maxim Khlopov
 Director of Virtual Institute of Astroparticle Physics,  for insightful comments and suggestions.

\end{document}